\documentclass[a4paper]{article}
\usepackage[margin=3cm]{geometry}
\usepackage{latexsym}
\usepackage{amsmath}
\usepackage{amsthm}
\usepackage{amssymb}
\usepackage{amsfonts}
\usepackage[pdftex]{graphicx}

\newtheorem{remark}{Remark}
\newtheorem{thm}{Theorem}
\begin{document}

\title{Parameter switching in a generalized Duffing system: Finding the stable attractors}
\author{Marius-F Danca$^{1,2}$ and Nicolae Lung$^3$ \\
$^{1}$ Department of Mathematics and Computer Science\\
Avram Iancu University\\
Str.\ Ilie Macelaru nr.\ 1A, 400380 Cluj-Napoca, Romania\\
$^{2}$ Romanian Institute of Science and Technology\\
Str. Ciresilor nr. 29, 400487 Cluj-Napoca, Romania \\
danca@rist.ro\\
$^{3}$  Dept. of Mathematics\\
Technical University of Cluj-Napoca\\
400114, Cluj-Napoca, Romania}

\maketitle

\begin{abstract}
This paper presents a simple periodic parameter-switching method which can find any stable limit cycle
that can be numerically approximated in a generalized Duffing system. In this method, the initial value
problem of the system is numerically integrated and the control parameter is switched periodically
within a chosen set of parameter values. The resulted attractor matches with the attractor
obtained by using the average of the switched values. The accurate match is verified by phase plots and
Hausdorff distance measure in extensive simulations.
\end{abstract}



%\noindent{\it PACS\/}: 02.60.Cb, 02.60.Jh, 05.45.Gg

\section{Introduction}\label{sectunu}
The well-known Duffing system was coined in 1918, which is one of the mostly studied nonlinear dynamical
systems describing mechanical structures, and electric circuits and even biological re systems.
This paper considers a generalized Duffing system of the form
\begin{equation}
\overset{..}{x}+a\overset{.}{x}+px+bx^{3}+c~sign(x)+d~sign(\overset{.}{x})=e~\cos
\left( \omega t\right), \label{unu}
\end{equation}
\noindent where $a, b, c, d, e, \omega$ and $p$ (considered as the control parameter) are real parameters.
As for almost all practical examples, at least one of the parameters, $c$ and $d$, will be zero. Thus,
according to different  functions of $c$ and $d$, one could have the classical form of excited Duffing
oscillator ($c = d = 0$), dry friction models ($c = 0, d = 1$), or other phenomena such as clearance,
vibro-impacts, and preloaded compliance ($c = 1, d = 0$). The external force is typically considered
to be periodic, since the study of the long-term behavior of an oscillator is relevant only in this setting.

Duffing's smooth and discontinuous dynamics are a very good examples for demonstrating how deterministic
chaos appears in mechanical systems that may be described as oscillators derived from a nonlinear potential.
For illustration, routes to chaos through bifurcations are shown in Fig.\ref{fig1}. There exists a large
volume of bibliography on the rich dynamics of the Duffing oscillator, some of the first titles being
\cite{guck,ueda} and \cite{parlitz}, while experimental implementations of the Duffing system can be found
in many references, e.g. \cite{holmes}.

Many non-smooth systems appear naturally in practical systems because such physical phenomena present
discontinuities, for instance the discontinuous dependence of friction force on the velocity, mechanical
structures under impacts and dry friction, brake processes with phase lock, oscillating systems with
combined dry and viscous damping, elasto-plasticity and forced vibrations. Also they appear in power
electrical circuits, convex optimization, control synthesis of uncertain systems, walking and hopping
robots, and even gene regulatory networks and neuronal networks, etc.
\cite{baner,bute,dei,polyn,popov,pop2,pop,solt,wir}. Noticeably, a large number of papers are devoted
to studying the fundamentals of discontinuous equations (such as (\ref{unu})) or to the afferent
differential inclusions which help tackle various difficult discontinuous problems
\cite{Aubin si Cellina,filip,kunze si kuper,kunze}. These studies clearly indicate that dry friction
and its underlying discontinuity present an important topic in both mathematical and engineering research.

Motivated by the above observations, considering the system (\ref{unu}) with discontinuity appears to be
a natural approach to more realistic engineering systems design and analysis. The present paper therefore
investigates an important and yet challenging problem in this system, more precisely a problem of
approximating (synthesizing) any stable attractor in system (\ref{unu}) by alternating parameter $p$
within a set of chosen values while the system is numerically integrated.

For this purpose, we will use the Parameter Switching (PS) algorithm. This algorithm is very effective
in approximating various complex dynamical behaviors corresponding to the switched parameter, such as
multiple attractors. It has been analytically proved \cite{Yu,Dan} that for a large class of continuous
systems, any synthesized attractor obtained by using this algorithm can well match with the attractor
obtained by replacing $p$ with the average of the alternated $p$ values. This has also been verified
numerically applicable to more general classes of dynamical systems. Moreover, the effectiveness of
the PS algorithm has been tested on several systems, including continuous, discontinuous, and fractional
or integer order systems \cite{danca1,dancakai}), such as Lorenz, R\"ossler, Chen, Chua, L\"u, Lotka
Volterra, and Hindmarsh-Rose neuronal systems, among others.\begin{footnote}{As shown in
\cite{danca1,dancakai,Yu} chaotic attractors can also be
synthesized. However, in this paper we are interested only in the
stable limit cycles.}\end{footnote}

System (\ref{unu}) is solved mathematically by the following general Initial Value Problem (IVP):
\begin{equation}
\overset{.}{x}=f(x)+pAx+Bs(x),~x(0)=x_{0},~t\in I=[0,\infty ).
\label{ivp generala}
\end{equation}
\noindent This shows that the system depends linearly on $p$, same as for the general class of many
known systems like the Lorenz, Chen, R\"{o}ssler, Chua,
Hindmarsh-Rose, Lotka Volterra systems. In (\ref{ivp generala}, $p \in \mathbb{R}$, $x\in
\mathbb{R}^{n}$, $A,B\in\Re^{n\times n}$ are constant matrices,
$~f:\mathbb{R}^{n}\rightarrow \mathbb{R} ^{n}$ is a nonlinear at
least continuous vector function, and $s:\mathbb{R}^{n}\rightarrow
\mathbb{R}^ {n}$ is a vector piecewise linear vector function being
composed of scalar signum functions, namely
\begin{equation*}
s(x)=\left(
\begin{array}{c}
sgn(x_{1}) \\
\vdots \\
sgn(x_{n})
\end{array}
\right).
\end{equation*}
Function of the entries of the matrix $B$, the IVP (\ref{ivp
generala}) can model continuous systems (when $B=O_{n\times n}$) or
discontinuous with respect to the state variable (Filippov like
systems \cite{filip}, when $B\neq O_{n\times n}$).

Now, consider the Duffing oscillator (\ref{unu}) in the phase space
$\mathbb{R}^3$ having the following three autonomous equations:
\begin{equation}
\begin{array}
{l}
\overset{.}{x}_{1}=  x_{2},\\
\overset{.}{x}_{2}= - a x_2-px_1-bx_1^{3}-c~sign(x_1)-d~sign(x_2)+e\cos \left(x_3\right),\label{IVP}\\
 \overset{.}{x}_{3}= \omega.
\end{array}
\end{equation}
\noindent It can be easily seen that this belongs to the class of systems described by the IVP
(\ref{ivp generala}) with
\begin{equation*}
\begin{array}{c}
f(x)=\left(
\begin{array}{c}
x_{2} \\
-ax_{2}-bx_{1}^{3}+e\cos \left( x_{3}\right)  \\
\omega%
\end{array}%
\right),~  A=\left(
\begin{array}{ccc}
0 & 0 & 0 \\
-1 & 0 & 0 \\
0 & 0 & 0%
\end{array}%
\right) ,~~B=\left(
\begin{array}{ccc}
0 & 0 & 0 \\
-c & -d & 0 \\
0 & 0 & 0%
\end{array}%
\right).
\end{array}%
\end{equation*}
Even the system (\ref{IVP}) is three-dimensional, so our interest here
is focused, on the phase plane $\left( x_1,x_2 \right)$, as usually for planar systems.

We will investigate the effect of the positive parameter $p$.\begin{footnote}{As is well known,
$p$ can also be negative (the "inverted" Duffing equation). Also, as known, any of the
coefficients $a,b,c,d,e$ or $\omega$ can be chosen as control
parameter.}\end{footnote} For the others parameters, we chose:

$a=1$ i.e. the case of a strong dissipation (damped oscillations),
in order to avoid long chaotic transients, typical for weak
dissipation (as known, chaotic behaviors could persist for some transient time before the
trajectory approach near the attractor \cite{yorke})\begin{footnote}{The effectiveness of
the PS algorithm is not influenced by the weak dissipation case, corresponding to
$a\ll$.}\end{footnote};

$b=1$;

$c$ and $d$ are chosen $0$ or $1$ corresponding to the continuous or
discontinuous case.

$e=37$ (the amplitude of the driving forces on oscillations $x$);

$\omega=0.88$.

\section{Attractors synthesis}

\subsection{\emph{Preliminary results and notions}}

\smallskip
\noindent \textbf{Notation} \textbf{1.} Let
$P_N=\{p_1,p_2,...,p_N\}$ a set of $N>0$ values of $p$. The
\emph{average value}, denoted by $p^*$ is given by
\begin{equation}
p^{\ast}=\frac{\sum\limits_{k=1}^{N}p_{k}m_{k}}{\sum\limits_{k=1}%
^{N}m_{k}}, \label{p*}
\end{equation}
\noindent where $m_i$ are some positive integers, which will be
precisely defined later.

\noindent \textbf{2.} We denote the attractors obtained through
alternating $p$ with the PS algorithm, the \emph{synthesized attractor},
by $A^*$ and the \emph{average attractor} by $A_{p^*}$, corresponding to
$p=p^*$.

\begin{remark}\label{pstea}
In different functions on $m_k$ values in (4), $p^*$ could be an element
of $\mathcal{P}_N$. However, in this
paper we consider that $p^*\notin \mathcal{P}_N$, since in practical
examples it is more realistic to approximate an attractor $A_{p^*}$
starting from a set $P_N$ which does not contain $p^*$.
\end{remark}

To understand how the PS algorithm works, we further consider
the general problem (\ref{ivp generala}) for the continuous case
($B=0_{n \times n}$), with a time-dependent $p$, as follows:
\begin{equation}
\dot{x}(t) = f(x(t)) + p\left(t\right) A x(t), ~~ x(t_0) = x_{0},
~~t \in I,\label{eq1}
\end{equation}
\noindent where $p:I\rightarrow P_N $ is considered a piecewise
constant periodic function with the period $T_0$, and the the mean
value $p^*$, namely,
\begin{equation*}
\frac{1}{ T_0}\int_t^{t+ T_0} p(u)du = p^*,  \qquad t\in I.
%%\label{eq2}
\end{equation*}
\noindent also, the \emph{average} model of (\ref{eq1}), is
expressed as follows;
\begin{equation}
\dot{y} = f(y) + p^* A y, ~~ y(0) = y_0. \label{eq2}
\end{equation}
\noindent Equation (\ref{eq1}) represents the mathematical model of
the PS algorithm.

\noindent In additiona, we need the following assumptions.

\noindent (\textbf{H1}) The IVP admits unique solutions (e.g., when $f$
is Lipschitz continuous).

\noindent (\textbf{H2}) To each $p$ value, there corresponds a single
attractor which will be numerically approximated by its
$\omega$-limit set \cite{Foias}, after neglecting a sufficiently
long period of transients.

\noindent (\textbf{H3}) The initial conditions $x_0$ and $y_0$ in
(\ref{eq1}) and (\ref{eq2}), respectively, are chosen close enough to
each other (in the same basin of attraction).

\vspace{3 mm}

\noindent Now, we can introduce the following theorem (proved in
$\mathbb{R}^n$ \cite {Yu,Dan}).

\begin{thm}\label{th}
The solution of Equation (\ref{eq1}) approaches the solution of Equation
(\ref{eq2}).
\end{thm}

\noindent The proof presented in \cite {Yu} is based on the averaging
theory \cite{sander}, and is done via generalized P\'{e}ano-Baker
series, while the proof presented in \cite{Dan} uses the convergence
of known numerical methods for ODEs.

\mbox{}

Thus, it is proved that the distance between the solutions of
linearized Equation (\ref{eq1}) and of Equation (\ref{eq2}), starting from the
same basin of attraction, is negligible. Therefore, we have revealed
(see also \cite{stuart}, Chapter 6) that the invariant sets of system
(\ref{eq1}), determined numerically, converge to the invariant sets
of system (\ref{eq2}). This means that the PS algorithm, modeled by
(\ref{eq1}), is approximated by periodical parameter switching and that
the attractor corresponding to $p^*$ is generated by (\ref{eq2}).

To Summarize, by switching $p$ periodically while the IVP is numerically
integrated, one obtains a synthesized attractor, $A^*$, which
matches with the attractor $A_{p^*}$ obtained when $p$ is replaced by
$p^*$.

The PS algorithm is useful in practical examples when one intends to
obtain some attractor $A^*$, but its underlying parameter $p^*$ cannot
be set. Thus, $p^*$ and the corresponding
attractor $A_{p^*}$ will be obtained by switching $p$ within some
accessible set of values $\mathcal{P}_N$.

\subsection{\emph{Numerical implementation}}

Theorem \ref{th} only proves that the PS algorithm convergences to some
attractor $A^*$, which approximates the attractor $A_{p^*}$, but it
does not indicate any way to implement it in concrete examples. Therefore,
a numerical modality to implement this result is necessary. For this
purpose, two steps are formulated:

\mbox{}

\noindent I) \emph{run the PS algorithm, which generates a
synthesized attractor $A^*$ via parameter switching;}

\noindent II) \emph{show numerically (aided by characteristic
tools for dynamical systems) that $A^*$ matches with the average
attractor $A_{p^*}$ obtained when $p$ is replaced by the average
value $p^*$.}

\begin{remark}
(i) Step II is necessary in order to prove that $A^*$ is not just an
attractor, but it belongs to the set of attractors of the underlying
system.

\noindent (ii) Due to the predominant numerical characteristics of
the present work, the time interval $I$ is considered hereafter
finite: $I=[0,T]$, with $T>0$.

\noindent (iii) Regarding the approach to the discontinuous case, it is
noted that the underlying IVP can be continuously approximated in
some neighborhood of the discontinuity point (here, $x_1=0$ and
$x_2=0$), using e.g. the Filippov regularization \cite{filip}. After
this, the PS algorithm is applied, as described for the continuous case
(see \cite{toti}). Thus, the problem is transformed to a continuous one,
where the PS algorithm is applicable.\begin{footnote}{One of the best
known books on the approximation theory of discontinuous IVP via differential
inclusions is \cite{Aubin si Cellina}.}\end{footnote}
\end{remark}

In order to reduce the number of transient steps and to
avoid possible complication when, for a given $p$
value, there are several (coexisting) attractors, the initial
conditions will be taken without loss of generality to be
$x_0=y_0$.

Let us again consider the simpler case of continuous Duffing
system ($B=0_{n \times n}$, i.e. $c=d=0$).

\vspace {3 mm}

I) To implement the PS algorithm, a numerical method for ODEs such
as the standard Runge-Kutta method with a fixed step seize $h$, will be
used. Suppose we chose $P_N$, and $p$ is switched indefinitely
within $P_N$ for $t<T$, in the following manner
\begin{equation*}
p(t)=p_i~~if~~t\in I_{i},~ p_i\in P_N, ~i=1,2,...,N,\\
\label{p}
\end{equation*}
\noindent where the time subintervals $I_i$, $i=1,2,...,N$, obtained
by partition of $I$, satisfy $I=\bigcup \left(\bigcup_{k=1}^{N}I_{k}\right)$.

The simplest way to realize that, numerically, is to choose the
length of $I_i$ as a multiple of $h$: $I_i=m_ih$, where $m_i,~
i=1,2,...,N$, are some positive integers ("weights").

Denote the PS algorithm, for a step size $h$, as follows
\begin{equation}
[m_1p_1,m_2p_2,...,m_Np_N].
\end{equation}

The pseudocode for PS algorithm is presented in Table \ref{tab1}.

For example, by $[1p_1,2p_2]$, we understand that PS, with $N=2$,
$P_2=\{p_1,p_2\}$, $m_1=1$ and $m_2=2$, integrates the IVP for one
step of size $h$ with $p=p_1$. Then, perform the next two steps with
$p=p_2$ and again one step with $p=p_1$; after that, perform two steps
$p=p_2$ and so on, until $t\geq T$, where the period $T_0=3h$ and
$p^*=(1\times p_1+2\times p_2)/(1+2)$.

If, for a given $P_N$ and a fixed $h$, we intend to obtain some $p^*$,
then we have to choose the set $m_1,m_2,...,m_N$, such that
(\ref{p*}) is verified. Reversely, it is possible to have the set
$P_N$ and the switching times $I_i$ (i.e., the set
\{$m_1,m_2,...,m_N$\} is given). Then, (\ref{p*}) will generate a
value for $p^*$.

\begin{remark}\label{pipi}
(i) As can be seen from relation (\ref{p*}), $p^*$ is a convex
combination of $p_k$. Therefore, $p^*$ will belong to the real open
interval $(p_1,p_N)$, if $p_k$, $k=1,2,...,N$, are considered to
be ordered. Hence, if we intend to generate some attractor
$A_{p^*}$, starting with the set $P_N=\{p_1,p_2,...,p_N\}$, a
necessary condition is that $p^*$, given by (\ref{p*}) satisfies
$p^*\in (p_1,p_N)$. However, this does not necessarily mean that
$p^*\in P_N$ (see Remark \ref{pstea}). Moreover, the convexity
implies that if $P_N$ is included in some periodic window, and
therefore contains only periodic values, then under whatever switching
scheme, the PS algorithm will lead to a stable periodic motion.

\noindent (ii) While the systems modeled by (\ref{ivp generala}) and
(\ref{eq2}) are autonomous ($p$ and $p^*$ are constant), (\ref{eq1})
models a nonautonomous system. Therefore, theoretically, the choice
of initial conditions depends on $t_0$. Let us consider, for
example, the scheme $[m_1p_1,m_2p_2]$. If $t_0$ belongs to interval
$I_1$, for which $p=p_1$, then the PS algorithm leads to the same
$A^*$, because the algorithm starts integration with $p_1$. The
result should be different if $t_0\in I_2$, for which $p=p_2$, when
the algorithm begins with $p_2$. However, after a number of transient
steps, the results show that $A^*$ does not depend on $t_0$.
Therefore, we can simply choose $t_0=0$.

\noindent(iii) It is easy to see that for a given set $P_N$,
Equation (\ref{p*}) has several sets of solutions $m_k$,
$k=1,2,...,N$. This means that choosing different schemes
$[m_1p_1,m_2,p_2,...,m_Np_N]$ with the same $P_N$ set, one can obtain
the same attractor $A^*$. Obviously, the same attractor $A^*$ can be
obtained with infinite many choices of $m_k$ and $P_N$ sets.
\label{maimulte}
\end{remark}

II) To numerically check that the synthesized attractor $A^*$
obtained with the PS algorithm matches with the average attractor
$A_{p^*}$, several tools can be used: superimposed histograms,
Poincar\'{e} sections, time series, phase plots and also Hausdorff
distance \cite{falco}, which is the most rigorous numerical match
verification (see Appendix). In this paper, we plot both attractors
$A^*$ and $A_{p^*}$ in the same phase plane and calculate their
Hausdorff distance to underline the match between them.

To summarize, using the PS algorithm one can do the following

\mbox{}

--synthesize any desired attractor corresponding to some
value $p^*$; for this purpose, we have to choose $N$, $P_N$ and
$m_1,m_2,...,m_N$, such that (\ref{p*}) is satisfied;

or

--choose $N$, $P_N$ and $m_1,m_2,...,m_N$, and apply the PS algorithm
to obtain some attractor $A^*$ (stable or chaotic), which belongs to
the set of all attractors of the considered system.

\mbox{}

Therefore, if we intend to find some attractor $A_p$ (stable limit
cycle here) for some value $p$, we have to choose $N$, $P_N$ and
the values $m_1,m_2,...,m_N$, such that (\ref{p*}) is satisfied when
$p^*$ is replaced by $p$. Then, applying the PS algorithm one obtains
$A^*$ which, as mentioned before, will be a numerical approximation
of the attractor $A_{p^*}$, i.e. the searched attractor.

\section{Finding stable limit cycles of the Duffing system}
\label{sectie}

\begin{figure*}
\begin{center}
\includegraphics[clip,width=1\textwidth]{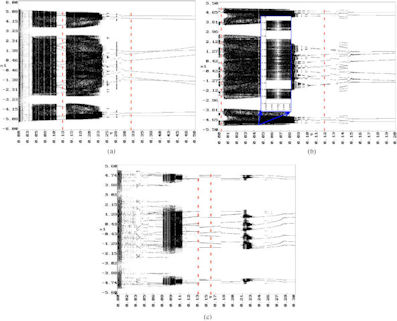}
\caption{}
\label{fig1}
\end{center}
\end{figure*}

The best way to study the effect of one specific parameter is to perform
bifurcation analysis with respect to it.
With the data presented in Section
\ref{sectunu}, the one-parametric bifurcation diagrams necessary for
both continuous and discontinuous cases are plotted in
Fig.\ref{fig1} a, b and c, respectively. As typical for most Duffing
type of systems, two types of routes to chaos can be found,
namely chaos after
(inverse) period doubling bifurcations (Feigenbaum route to chaos)
and the intermittent type (arising at the edge of
Feigenbaum bifurcation). Also, sudden changes in the size of a chaotic
attractor and in the number of unstable periodic orbits (crisis) can
be viewed in all three bifurcation diagrams shown in the figure.
For the discontinuous
case, one can observe a typical abruptly stability change
(possible hysteresis) (Fig.\ref{fig1} b and c).

All the numerical tests have been performed with the standard Runge
Kutta scheme with, unless specified otherwise, $h=0.005$, $T=500$
and initial
conditions $(0.1, 0.1, 0.1)$. The results are summarized in
Table \ref{tabelmare}.

As is well known, the Duffing system presents strong asymptotic behavior.
Therefore, as stated above, the beginning transient steps are
neglected. The used values for $p^*$ are plotted with dashed lines
in the bifurcation diagrams in the above figures.
The calculated Hausdorff distance,
$D_H$, with a few exceptions (related to the PS algorithm limits, see Section
\ref{limit}), is of order $10^{-3}$, which confirms a
good approximation. Supplementarily, to verify the match between
$A^*$ and $A_{p^*}$, beside $D_H$, both attractors are plotted
superimposedly in the phase plane (in blue and red, respectively).

\subsection{A. Continuous case of $c=d=0$}\label{sec}

Consider the IVP (\ref{IVP}) with $B=O_{3\times 3},
~(c=d=0$):
\begin{equation}
\begin{array}
[c]{cl}
\overset{.}{x}_{1}= & x_{2},\\
\overset{.}{x}_{2}= & - x_2-px_1-x_1^{3}+37\cos \left(x_3
\right),\label{cont}\\
 \overset{.}{x}_{3}=& 0.88.
\end{array}
\end{equation}

\begin{figure*}
\includegraphics[clip,width=1\textwidth]{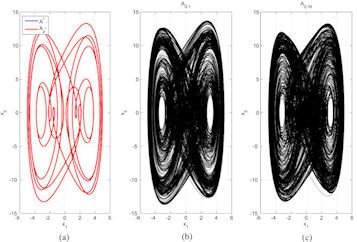}
\caption{} \label{fig2}
\end{figure*}

\noindent (a) \noindent Suppose we intend to obtain a stable higher-periodic
limit cycle corresponding to $p=0.13$ (see Fig. \ref{fig1}) by
using $N=2$ values for $p$: $P_2=\{0.1,0.16\}$. This means that in
(\ref{p*}), we replace $p^*$ with $0.13$ and find one of the possible
solutions for $m_k$ (see Remark \ref{maimulte} (iii)), e.g.
$m_1=m_2=1$. With these values, the PS algorithm can then be applied
to obtain $A^*$, which is the numerical approximation of $A_p^*$
with $p^*=0.13$. The attractors corresponding to $p=p_1$ and $p=p_2$
($A_{0.1}$ and $A_{0.16}$ respectively) are chaotic (see
projections of the phase portraits in Fig. \ref{fig2} b and c), while
the synthesized and average attractors $A^*$ and $A_{p^*}$, with
$p^*=0.13$, are indeed stable higher-periodic cycles (Fig.\ref{fig2}a).

\mbox{}

\begin{figure}
\includegraphics[clip,width=1\textwidth]{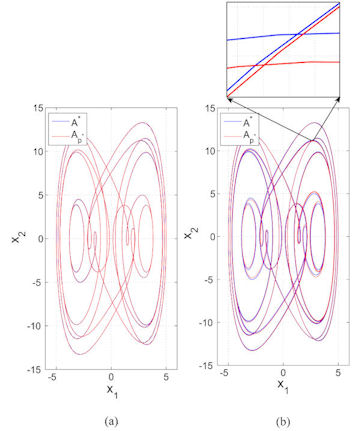}
\caption{} \label{fig3}
\end{figure}

\noindent (b) The same attractor (Remark \ref{maimulte} (iii))
$A^*$, with $p^*=0.13$, can be obtained, e.g. with $N=4$, using the
scheme $[2p_1,1p_2,1p_3,2p_4]$, for $p_1=0.11,~p_2=0.12,~p_3=0.14,~
p_4=0.15$ (Fig. \ref{fig3}a).

\mbox{}

\noindent (c) As shown above, an rather arbitrary attractor
can be obtained with a larger number $N$.
For example, $A_{0.13}$ can be synthesized
with $N=21$ and $p_k=0.05+k\times 0.01, k=1,2,...,20, k\neq 8$ (see
Remark \ref{pstea}) and $m_1=3, m_2=4, m_3=2, m_4=4,
m_5=m_6=m_7=m_8=m_9=1, m_{10}=m_{11}=2, m_{12}=...=m_{21}=1$
(Fig. \ref{fig3}b).

\mbox{}

\noindent (d) Stable limit cycles can be obtained even if $P_N$
contains only periodic values, e.g. embedded in a periodic window
(see Remark \ref{pipi} (i)). For example, with $P_N=\{0.27,0.49\}$
and the scheme $[3p_1,1p_2]$, one obtains the stable limit cycle
$A_{0.325}$. In Fig. \ref{fig4} a, all the attractors are plotted in
the same phase plane, so as to compare $A^*$ with the
underlying attractors $A_{0.27}$ and $A_{0.49}$.

\mbox{}

\noindent (e) As shown in \cite{haos}, the PS algorithm can be
applied in a certain random manner. In so doing, the $p$
values will not be alternated within $P_N$ in a periodic
(deterministic) manner, but rather randomly. However, in
this case one obtains an average value $p^o$, which is only
approximatively close to $p^*$ and has to be determined with the
following formula:
\begin{equation*}
p^{o}=\frac{\sum\limits_{k=1}^{N}p_{k}m'_{k}}{\sum\limits_{k=1}%
^{N}m'_{k}}, \label{po}
\end{equation*}
\noindent where $m'_k$ counts the number of $p_k$ during the
integration over $I$.

For example, if one chooses $N=2$ and switch $p$ randomly (with uniform
distribution) within the set $\{0.12,0.14\}$, then after $200000$
steps with $h=0.005$ and $p^o=0.13001$, the attractor $A^o$ is still
close to $A_{p^*}$. However, some difference between $A^*$ and
$A^*$, like those shown in Fig. \ref{fig3} b, can be observed
(Fig. \ref{fig4} b).
In this case, $D_H$ is only of order $10^{-2}$.

\begin{remark}
(i) Obviously, for the value $p^o$ to be closer to $p^*$, the
integration time interval $I=[0,T]$ has to be larger than that for
deterministic switching. However, we cannot expect that an
asymptotic increase of $I$ (i.e. $T\rightarrow \infty$) will finally
imply $p^o=p^*$, since the global error for a convergent method (like
the Runge-Kutta scheme used here) grows exponentially. For example,
for the Runge-Kutta method, the global error is \cite{stuart}
$K/Lh^r(e^{LT}-1)$, where $L$ is the Lipschitz constant of the right-hand
side of the corresponding IVP, $r$ is the method order, and $K$ is
some constant. Thus, the global error depends exponentially on the size
$T$. Nevertheless, in our numerical experiments, for random switching,
with reasonable $T$ values of order $10^3$ (e.g. $T=1000$, i.e.
more than twice of that for the periodic case), we obtain
$||p^*-p^o||<10^{-5}$.

\noindent (ii) Now, it becomes obvious that the above-mentioned
periodicity of $p$ is not a necessary condition.

\end{remark}

\subsection{B. Discontinuous case of $c=0$ and $d=1$}

In this case, the system becomes
\begin{equation}
\begin{array}
[c]{cl}
\overset{.}{x}_{1}= & x_{2},\\
\overset{.}{x}_{2}= & - x_2-px_1-x_1^{3}-sign(x_2)+37\cos \left(x_3
\right),\label{discont1}\\
 \overset{.}{x}_{3}=& 0.88.
\end{array}
\end{equation}
\noindent As mentioned above, the discontinuous IVP can be
continuously approximated in a small neighborhood of $(x_1,0,x_3)$,
after which the PS algorithm can be applied.

\mbox{}

\noindent (a) To obtain a stable limit cycle, corresponding to e.g.
$p=0.0375$, we can use the scheme $[1p_1,1p_2]$ with $p_1=0$ and
$p_2=0.075$ (Fig. \ref{fig5} a), for which $p^*=0.0375$.

\mbox{}

\noindent (b) With $N=10$ and scheme $[m_1p_1,...,m_{10}p_{10}]$,
$p_k=0.05+k~0.01$, $k=1,...,11$, $k\neq 6$ (see Remark \ref{pstea})
and $m_1=...=m_9=1$, $m_{10}$=2, another stable limit cycle
corresponding to $p=0.12$ can be obtained (see Fig. \ref{fig1}
b). The attractors $A^*$ and $A_{p^*}$ are plotted in Fig.\ref{fig5}
b.

As can be seen from Fig.\ref{fig1} b, there exists an apparently periodic
window corresponding to $p\simeq 0.045$, which actually is a chaotic
window.

\subsection{C. Discontinuous case of $c=1$ and $d=0$}

With $c=1$ and $d=0$, the system has the following form
\begin{equation}
\begin{array}
[c]{cl}
\overset{.}{x}_{1}= & x_{2},\\
\overset{.}{x}_{2}= & -  x_2-px_1- x_1^{3}-sign(x_1)+37\cos
\left(x_3
\right),\label{discont2}\\
 \overset{.}{x}_{3}=& 0.88.
\end{array}
\end{equation}

The discontinuous IVP is continuously approximated as did
in Subsection 3.2.

\noindent (a) Consider the stable limit cycle $A_{0.16}$
(Fig. \ref{fig1} c). This stable attractor can be obtained with
scheme $[1\times 0.1,1\times 0.22]$ . The attractors $A^*$ and
$A_{p^*}$ are plotted superimposedly in Fig. \ref{fig6} a, while
$A_{0.1}$ and $A_{0.22}$ are plotted in Fig. \ref{fig6} b and c,
respectively.

\mbox{}

\noindent (b) To obtain another stable limit cycle, $A_{0.135}$ (see
Fig. \ref{fig1} c) with $N=10$, we use the scheme
$[m_1,m_2,...,m_{10}p_{10}]$ with $p_k=(k-1)0.03$, and $m_k=1$ for
$k=1,2,...,10$. Attractors $A^*$ and $A_{p^*}$, with $p^*=0.135$,
are plotted superimposedly in Fig. \ref{fig6} d.

\section{PS algorithm limits}\label{limit}

As expected, the numerical PS algorithm has performance
limits due to several factors, such as: errors of the
numerical method, lengths of the time-subintervals $I_k$,
$k=1,2,...,N$, i.e. sizes of $m_k$, the $N$ value, the digit number of
$p$, the step size $h$, and the distance in the parameter space
between different $p_k$. Also, the way in which $p$ is switched
(deterministic or randomly) is another factor that influences the
PS algorithm performances.

We now present more precise discussions on this concerned issue.

\mbox{}

\noindent \emph{Influence of $N$}

\noindent Actually, $N$ is not an influential factor if the step size
$h$ is chosen to be small enough. Thus, $N$ can even be of order $10^2$
without influencing substantially the accuracy of the results.

\mbox{}

\noindent \emph{Influence of the $I_k$ length}

\noindent This parameter measures the ``weight" of each $p_k$
value. It is a critical parameter. We consider here the case of
discontinuous Duffing system (\ref{discont1}) with $N=2$,
$P_N=\{0.12,0.14\}$, $T=500$, and the scheme $[m_1p_1,m_2p_2]$.
Here, $m_1$
and $m_2$ will be chosen equal, such that $p^*=0.13$ for all
considered examples. It is obvious that large $I_k$ (or $m_k$)
may influence the convergence of $A^*$ to $A_{p^*}$. Its influence
should be considered together with that of $h$. For example, if we
consider $h=0.005$, a superior limit for $m_1$ and $m_2$ could be
$25$, i.e. length $I_k=25h$, since $A^*$ and $A_{p^*}$ do not match
properly (see Fig. \ref{fig7} a). However, for a smaller step size
$h=0.001$, the difference diminishes (Fig. \ref{fig7} b). If we
consider a larger value for $N$, e.g. $N=35$, then $h=0.005$ is no
longer a suitable value (Fig. \ref{fig7} c) and this happens even if a
smaller value is chosen for $h$, e.g. $h=0.001$ (Fig. \ref{fig7} d).

\mbox{}
\noindent \emph{Influence of the $h$ size}

\noindent The $h$ value is another important factor that influences the
results together with $m_k$, as shown above. In our examples, $h$
should be taken to be about $0.005$. Whatever are the $m_k$ values, larger
step sizes can lead to mismatches especially because of the errors
induced by the used numerical method, while smaller values of
$h$ (e.g. of order $10^{-4}$) could be considered, but at the cost
of the computational time, which has no significant increase of
accuracy of the PS algorithm.

\mbox{}

\noindent \emph{Influence of the distance between different $p_k$}

\noindent As can be seen in the examples considered above, this parameter
in the PS algorithm does not influence the performances.

\mbox{}

\noindent \emph{Influence of the $p$ digits}

\noindent Another source of errors is the accuracy in presenting
the value $p$. Even though the program codes we use can deal with a
high but finite precision, it is not helpful to use more than 4 decimals
for $p^*$. For example, the width of some (periodic or chaotic) windows
in the parametric space is of order $10^{-4}$ (see
Fig. \ref{fig1} b).

\section{Conclusions and Discussions}

In this paper, we have shown numerically that any stable attractor
(limit cycle) of a generalized Duffing system can be well approximated
by simple parametric switching, with main results summarized in Table
\ref{tabelmare}. The switching can be performed in either some
deterministic way or random manner within a specified set of values.
The only necessary condition is that the targeted value of parameter $p$,
being replaced in (\ref{p*}), is located inside the real open interval
$(p_1,p_N)$, due to the convex property of the set of $p^*$ values.

Using the PS algorithm, not only regular but also chaotic motions can
be well approximated. Therefore, the PS algorithm can be viewed as a
kind of control/anticontrol algorithm \cite{danca1}, which can be
used whenever some targeted value $p^*$ cannot be accessed directly due
to some technical reasons. Compared to the classical control/anticontrol
methods, where e.g. an unstable periodic orbit (UPO) is transformed into
a stable one, here we synthesize an already existing stable orbit. Also,
one of the most important and useful features of the PS algorithm is
that the differences between the $p_k$ values can be arbitrarily
large in contrast to the classical control/anticontrol schemes.

The PS algorithm can be used to explain why in some real systems,
accident switching of a parameter could significantly change the
behavior of the system. It can also be used to illustrate how to obtain
a desired behavior starting from an
accessible set of parameter values.

How to implement experimentally the PS algorithm into real
systems should be investigated. From the existing possibilities,
we may choose the schemes $[m_1p_1,...,m_Np_N]$, for fixed $N$,
which are the ones with large time intervals $I_k$ (high values $m_k$).

For small differences between different elements of $P_N$, with $N$
sufficiently large, we could consider that the PS algorithm acts like
inducing some kind of parametric noise. Thus, in this case, by involving
parametric noise, we can find transition from a stable or unstable
state to another stable state.

\begin{figure}
\includegraphics[clip,width=1\textwidth]{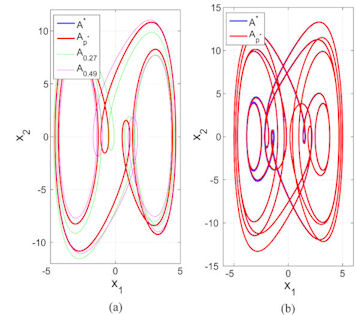}
\caption{} \label{fig4}
\end{figure}

\begin{figure}
\includegraphics[clip,width=1\textwidth]{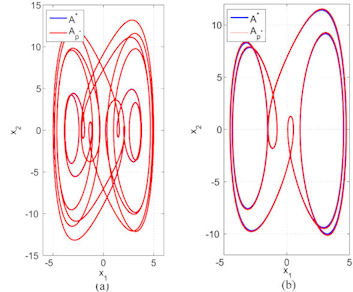}
\caption{} \label{fig5}
\end{figure}

\begin{figure}
\includegraphics[clip,width=1\textwidth]{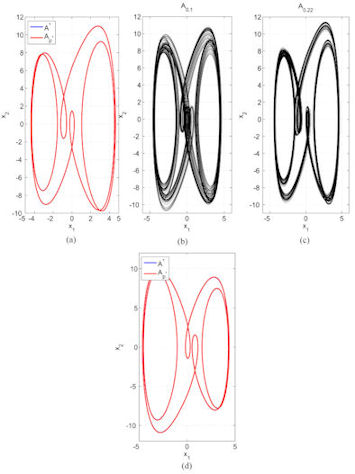}
\caption{} \label{fig6}
\end{figure}

\begin{figure}
\includegraphics[clip,width=0.7\textwidth]{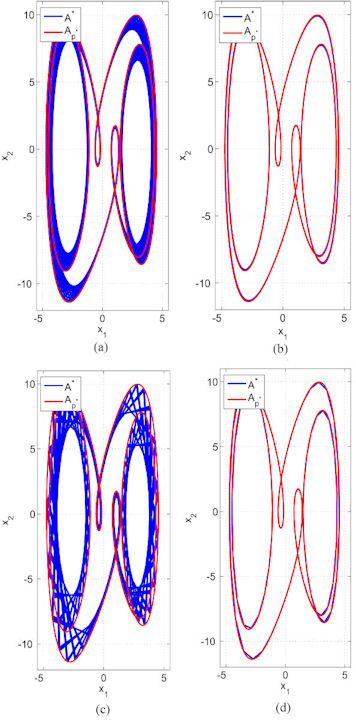}
\caption{}\label{fig7}
\end{figure}

The existence of an isomorphism between $P_N$ and the set of all
attractors of the system (belonging to the class of considered
systems) could be useful to show that, following the convex
property of $p^*$, $A^*$ might be a kind of
``convex combination" of the attractors $A_{p_1},
A_{p_2},...,A_{p_N}$, in the state space (see, e.g.,
Fig. \ref{fig4} a).

\begin{figure}
\includegraphics[clip,width=1\textwidth]{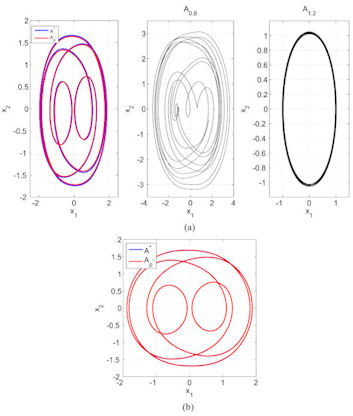}
\caption{}\label{fig88}
\end{figure}

The PS algorithm seems to work for other more general classes of
systems, not only for $p$-linear systems modeled by (\ref{IVP}).
Thus, we may consider the archetypal oscillator
\cite{wierci2}, which bears significant similarities to the Duffing
oscillator, given by
\begin{equation}
\overset{..}{x}+2\xi \overset{.}{x}+x\left(
1-\frac{1}{\sqrt{x^{2}+\alpha ^{2}}}\right) =f_{0}\cos \left( \omega
t \right), \label{w1}
\end{equation}
\noindent with the control parameter $p=\alpha$ and the other
parameters particularized as follows:
\begin{equation*}%
\begin{array}
[c]{cl}%
\overset{\cdot}{x}_{1}= & x_{2},\\
\overset{\cdot}{x}_{2}= & -0.0282x_{2}-x_{1}\left (1-\frac{1}{\sqrt{x_{1}%
^{2}+p^{2}}  }\right)+0.8cos(x_{3}),\\
\overset{\cdot}{x}_{3}= & 1.0607.
\end{array}
\end{equation*}
\noindent By using the PS algorithm with $N=2$, $P_N=\{0.8, 1.2\}$ and
scheme $[1m_1p_1,1m_2p_2]$, a similarity between $A^*$ and $A_{p^*}$
with $p^*=1$, can still be recognized, although the attractors do not
match as well as for the case of (\ref{IVP}) (see Fig. \ref{fig88} a).
Precisely, the relation (\ref{p*}) does not hold.

However, if we consider $\xi$ as the control parameter, the system
(\ref{w1}) belongs to the class of systems modeled by (\ref{ivp
generala}) and the PS algorithm can still be applied even with $N=100$
values which, for $p_k=k\times0.0002$ and $m_k=1$, $k=1,2,...,100$,
yields $p^*=0.0101$ (Fig. \ref{fig88} b).

\vspace{3mm}

\textbf{Acknowledgement} The authors acknowledge useful discussions
with Professor Marian Wiercigroch.

\pagebreak

\textbf{Appendix} Hausdorff distance between two sets \vspace{3 mm}

Consider a metric space. As is well known, in order to calculate
Euclidean distance between two sets, we have to find some
Euclidean isometry such that they become aligned, a difficult task in
our present study. This inconvenience can be avoided if we use Hausdorff
distance instead,
which looks only at the interpoint distance between the points on
each set.

The Hausdorff distance (or Hausdorff metric) $D_H$ measures how far
two compact nonempty subsets of the considered metric space are
from each other. Since the considered attractors (stable
limit cycles here) are nonempty compact sets, we can calculate
$D_H$. Here, two sets are close to each other in the Hausdorff
distance if every element of a set is close to some element of
the other set.

\begin{figure}
  \includegraphics[clip,width=1\textwidth] {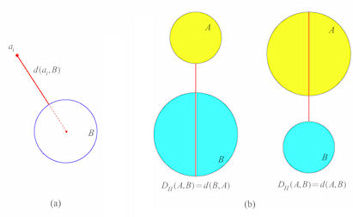}
  \caption{}\label{dh}
\end{figure}

The Hausdorff distance between two curves in $\mathbb{R}^n$ is
defined as the maximum distance to the closest point between the
curves. If the curves are defined, as in our case, as the sets of
ordered pair of coordinates $A=\{a_1, a_2,..., a_{k_1}\}$, $B=\{b_1,
b_2,...,b_{k_2}\}$, with $a_i=(x_1,x_2,...,x_n)$ and
$b_j=(y_1,y_2,...,y_n)$, then $D_H$ can be expressed as follows
(Fig. \ref{dh}):
\begin{equation}
D_{H}\left( A,B\right) =\max \left\{ d\left( A,B\right),
~d(B,A)\right\}, \label{haus}
\end{equation}
\noindent where the distance between $A$ and $B$, denoted by $d(A,B)$
(generally different from $d(B,A)$), has the following form:
\begin{equation*}
d(A,B)=\underset{i}{\max }\left\{ d\left( a_{i},B\right) \right\},
\end{equation*}
\noindent and is defined via the Euclidean distance between $a_i$
and $B$ (Fig. \ref{dh}a) as
\begin{equation*}
d\left( a_{i},B\right) =\underset{j}{\min }||a_{i}-b_{j}||.
\end{equation*}
Compared with other conventional methods, which require substantial
computing time, $D_H$ is very easy to calculated numerically. The
only requirement to apply the relation (\ref{haus}), e.g. for our examples,
is the number of points on each curve ($k_1$ and $k_2$ respectively) must be
large enough, so as to well describe the entire curve.

\newpage

\clearpage

\clearpage

\clearpage

\clearpage

\clearpage

\clearpage

\clearpage

\clearpage

\newpage

\begin{table}
\[%
\begin{array}
[c]{l}%
\hline
Input:~N,~P_N,~T,~h,~m_{1},\ldots,m_{N}\\
t=0\\
Repeat\\
~~~~~~For~k=1~to~N~do\\
\quad~~~\quad~~~~p=p_{k}\\
~~~~~~~~~~~~~\,for~i=1~to~m_{k}~do\\
~~~\ ~~~~~~~~~~~~~~~~integrate~IVP~(\ref{ivp generala})\\
~~~~~~~~~~~~~~~~~~~~t=t+h\\
~~~~~~~~~~~~~end\\
~~~~~~~end\\
until~t\geq T_{\max}\\
\hline
\end{array}
\]
\caption{}\label{tab1}
\end{table}

\begin{table*}
\newcommand\T{\rule{0pt}{2.6ex}}
\newcommand\B{\rule[-1.2ex]{0pt}{0pt}}
\begin{center}
\begin{tabular}{|c|c|c|c|c|c|}
\hline
System & Scheme & $m$ & $P_{N}$ & $p^*$ & Fig.  \\
\hline& \lbrack $m_{1}p_{1},m_{2}p_{2}]$ & $m_{1}=m_{2}=1$ &
$p_{1}=0.1,~p_{2}=0.16$
& $0.13$ &  \ref{fig2}\\
\cline{2-6} & \lbrack $m_{1}p_{1},...,m_{4}p_{4}]$ & $
m_{1}=m_{4}=2,~m_{2}=m_{3}=1$ & $%
\begin{array}{c}
p_{1}=0.11,~p_{2}=0.12, \\
~p_{3}=0.14,~p_{4}=0.15
\end{array}
$ & $0.13$ & \ref{fig3}a \\
\cline{2-6}$
\begin{array}{c}
IVP (\ref{cont})\\
c=0\\
d=0
\end{array}
$&\lbrack $m_{1}p_{1},...,m_{21}p_{21}]$ & $%
\begin{array}{c}
m_{1}=3,~m_{2}=m_4=4,~m_{3}=2, \\
m_{5}=...=m_{8}=1 \\
m_{9}=m_{10}=2, \\
m_{11}=...=m_{21}=1
\end{array}
$ & $
\begin{array}{c}
p_{k}=0.05+k\times 0.01, \\
k=0,1,...,21,~k\neq 8
\end{array}
$ & $0.13$ & \ref{fig3} b \\
\cline{2-6}
& $\lbrack m_{1}p_{1},m_{2}p_{2}]$ & $m_{1}=3,~m_{2}=1$ & $%
p_{1}=0.27,~p_{2}=0.49$ & $0.325$& \ref{fig4} a\\
\cline{2-6}
& random scheme &  & $%
p_{1}=0.12,~p_{2}=0.14$ &$%
\begin{array}{c}
p^o=0.13001\\
p^*=0.13
\end{array}
$& \ref{fig4} b\\
\hline & $\lbrack m_{1}p_{1},m_{2}p_{2}]$ & $
\begin{array}{c}
m_{1}=1,~m_{2}=1 \\
\end{array}
$ & $
\begin{array}{c}
p_{1}=0, p_2=0.075
\end{array}
$ & $0.0375$ & \ref{fig5} a \\
\cline{2-6} $
\begin{array}{c}
IVP (\ref{discont1})\\
c=0\\
d=1
\end{array}
$ & \lbrack $m_{1}p_{1},...,m_{10}p_{10}]$ & $
\begin{array}{c}
m_{1}=...=m_9=1,  \\
m_{10}=2, \\
\end{array}
$ & $
\begin{array}{c}
p_{k}=0.05+k\times 0.01, \\
k=1,2,...,11,~k\neq 6
\end{array}
$ & $0.12$ &  \ref{fig5} b\\
\hline   $
\begin{array}{c}
\\IVP (\ref{discont2})\\
c=1\\
d=0
\end{array}
$& \lbrack $m_{1}p_{1},m_{2}p_{2}]$ & $
\begin{array}{c}
m_{1}=1,~m_{2}=1
\end{array}$ & $
\begin{array}{c}
p_{1}=0.04, p_2=0.28
\end{array}
$ & $0.16$ &  \ref{fig6} a\\
\cline{2-6} & $\lbrack m_{1}p_{1},...,m_{10}p_{10}]$ & $
\begin{array}{c}
m_{1}=...=m_{10}=1 \\
\end{array}
$ & $
\begin{array}{c}
p_{k}=(k-1)\times0.03,\\k=1,...,10
\end{array}
$ & $0.135$ & \ref{fig6} b  \\
\hline
\end{tabular}
\caption{}
\label{tabelmare}
\end{center}
\end{table*}

\clearpage
Figure captions

Fig. 1. Bifurcation diagram of the Duffing system (\ref{IVP}). The
dashed lines present the parameter values corresponding to the
synthesized attractors. (a) Continuous case ($c=d=0$). (b)
Discontinuous case ($c=0,~ d=1$). (c) Discontinuous case ($c=1,~ d=0$).

Fig. 2. The PS algorithm applied to the continuous Duffing
system (\ref{cont}) with $N=2$, $P_2=\{0.1,0.16\}$ and $m_1=m_2=1$.
(a) $A^*$ and $A_{p^*}$, with $p^*=0.13$. (b) Attractor $A_{0.1}$.
(c) Attractor $A_{0.16}$.

Fig. 3. The stable limit cycle $A_{0.13}$ of the continuous Duffing
system (\ref{cont}) obtained with $\left( a \right )$
$[2p_1,1p_2,1p_3,2p_4]$, for $p_1=0.11, p_2=0.12, p_3=0.14,
p_4=0.15$. (b) Same attractor $A_{0.13}$ for $N=21$ with
$p_k=0.05+k\times 0.01, k=1,2,...,20, k\neq 8$ and $m_1=3, m_2=4,
m_3=2, m_4=4, m_5=m_6=m_7=m_8=m_9=1, m_{10}=m_{11}=2,
m_{12}=...=m_{21}=1$. Both attractors $A^*$ and $A_{p^*}$, with
$p^*=0.13$, are plotted superimposedly.

Fig. 4. (a) The stable limit cycle $A_{0.325}$ of the continuous
Duffing system (\ref{cont}) obtained with $P_N=\{0.27,0.49\}$ and
the scheme $[3p_1,1p_2\}$. All the attractors, $A^*$, $A_{p^*}$ (with
$p^*=0.325$), $A_{0.27}$, and $A_{0.49}$, are plotted in the same
phase plane. (b) The PS algorithm applied with uniformly distributed
random switching of $p$ within the set $\{0.12,0.14\}$ to obtain the
attractor $A_{0.13}$.

Fig. 5. (a) Stable limit cycle $A_{0.0375}$ for the discontinuous
Duffing system (\ref{discont1}), obtained with $[m_1p_1,m_2p_2]$ for
$p1=0, p2=0.075$ and $m_1=m_2=1$. b) Stable limit cycle $A_{0.12}$
obtained with the scheme $[m_1p_1,...,m_{10}p_{10}]$, with
$p_k=k0.01+0.05$, $k=1,...,11$, $k\neq 6$ and $m_1=...=m_9=1$,
$m_{10}$=2.

Fig. 6. Stable limit cycle $A_{0.16}$ of the discontinuous Duffing
system (\ref{discont2}) obtained with the scheme $[1\times
0.1,1\times 0.22]$. (a) $A^*$ and $A_{p^*}$ with $p^*=0.16$. (b)
$A_{0.1}$. (c) $A_{0.22}$. (d) Stable limit cycle $A_{0.135}$
obtained with the scheme $[m_1p_1,m_2p_2,...,m_{10}p_{10}]$ with
$p_k=(i-1)0.03$, $i=1,2,...,10$ and $m_k=1$, $k=1,2,...,10$.
$p^*=0.135$.

Fig. 7. Stable limit cycle $A_{0.13}$ of the discontinuous Duffing
system (\ref{discont1}) obtained with $[m_1p_1,m_2p_2]$ with: (a)
$m_1=m_2=25$ and $h=0.005$. (b) $m_1=m_2=25$, $h=0.001$. (c)
$m_1=m_2=35$, $h=0.005$. (d) $m_1=m_2=35$, $h=0.001$.

Fig. 8. (a) The PS algorithm applied to the system (\ref{w1}) for
$p=\alpha$ as control parameter and with $N=2$, $P_N=\{0.8, 1.2\}$
and scheme $[1p_1,p_2]$. (b) The PS algorithm with $N=100$, applied to
the same system, but with $p=\xi$.

Fig. 9. Hausdorff hdistance. (a) $d(a_i,B)$; (b) $D_H$ for two ideal cases.
A suggested applet can be found from \cite{inter}.

\clearpage
Table captions

Table 1. Pseudo-code of the PS algorithm.

Table 2. The PS algorithm applied to the Duffing systems
(\ref{cont}), (\ref{discont1}) and (\ref{discont2}).

\end{document}